# Minimum Grounded Component Based Voltage-Mode Quadrature Oscillator using DVCC

J. Mohan[1], S. Maheshwari[2], and D. S. Chauhan[3]
[1]Department of Electronics and Communications, Jaypee University of Information Technology,
Waknaghat, Solan-173215 (India)
Email:jitendramv2000@rediffmail.com
[2]Department of Electronics Engineering, Z. H. College of Engineering and Technology,
Aligarh Muslim University, Aligarh-202002 (India)
Email:sudhanshu_maheshwari@rediffmail.com
[3]Department of Electrical Engineering, Institute of Technology, Banaras Hindu University,
Varanasi-221005 (India)
Email:pdschauhan@gmail.com

*Abstract*— **In this paper, a new voltage-mode quadrature oscillator using minimum number of active and passive component is proposed. The proposed circuit employs single modified DVCC, two grounded capacitor and two grounded resistors, which is ideal for IC implementation. The active and passive sensitivity are no more than unity. The proposed circuit is verified through PSPICE simulation results.**

## I. INTRODUCTION

Quadrature sinusoidal oscillator produces two sinusoidal outputs of identical frequency but of 90° phase shift. It is used for a wide variety of applications such as in telecommunication for quadrature mixers, in single side band generators, indirect conversion receivers or for measurement purposes in vector generator or selective voltmeters [1-2]. As a result, many quadrature oscillator circuits have been reported in the literature [3-17] using different type of active elements. The circuit configuration presented in [17] uses four grounded passive components and single active element, but uses a capacitor at terminal X, which degrades the high frequency operation [18].

This paper presents a new voltage-mode quadrature oscillator (VMQO) using minimum number of active and passive grounded component. The proposed circuit employs single modified DVCC, two grounded capacitor, two and grounded resistors. The active element used is DVCC which has been recently used for analogue signal processing by several researchers [19-20]. The use of grounded capacitors and resistors make the proposed circuit suitable for integrated circuit implementation [21]. PSPICE simulation results are given to validate the new proposed circuit.

## II. PROPOSED CIRCUIT

A differential voltage current conveyor (DVCC) was first introduced by Pal [22] and further developed by Elwan and Soliman [23]. An early application of the active element was reported by Gupta and Senani [7]. The basic aim of introducing DVCC was to add the differential input voltage capability to the already versatile second generation current conveyor [24-25]. The symbol and CMOS implementation of the modified DVCC are shown in Fig. 1. The modified DVCC can be characterized by

$$\begin{bmatrix} V_X \\ I_{Y1} \\ I_{Y2} \\ I_{Z1+} \\ I_{Z2+} \end{bmatrix} = \begin{bmatrix} 0 & 1 & -1 & 0 \\ 0 & 0 & 0 & 0 \\ 0 & 0 & 0 & 0 \\ 1 & 0 & 0 & 0 \\ 1 & 0 & 0 & 0 \end{bmatrix} \begin{bmatrix} I_X \\ V_{Y1} \\ V_{Y2} \\ V_Z \end{bmatrix} \quad (1)$$

The proposed voltage mode quadrature oscillator (VMQO) is shown in Fig. 2. It is composed of single modified DVCC, two grounded capacitors and two grounded resistors. The characteristic equation of the circuit is expressed as below

$$s^2 + s\left(\frac{1}{R_1 C_1} + \frac{1}{R_2 C_2} - \frac{1}{C_1 R_2}\right) + \frac{1}{R_1 R_2 C_1 C_2} = 0 \quad (2)$$

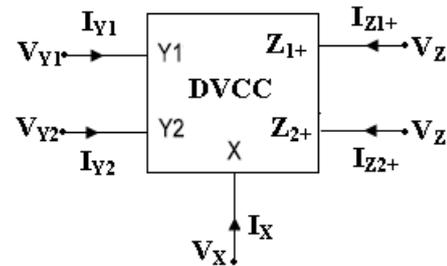

(a)

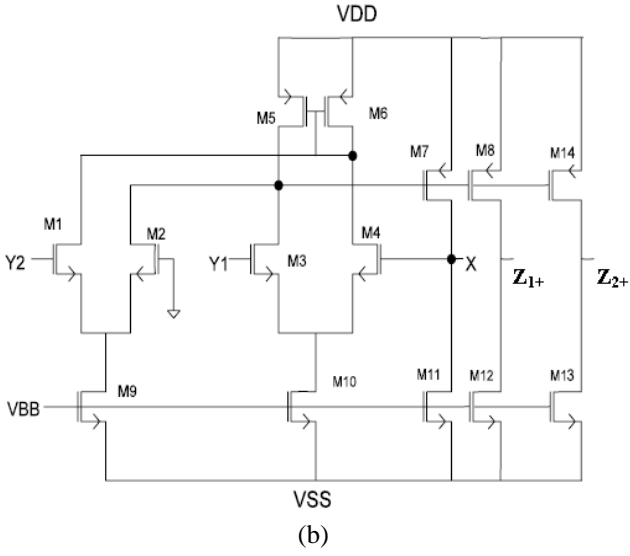

(b)

Figure 1. (a) Symbol (b) CMOS implementation of Modified DVCC

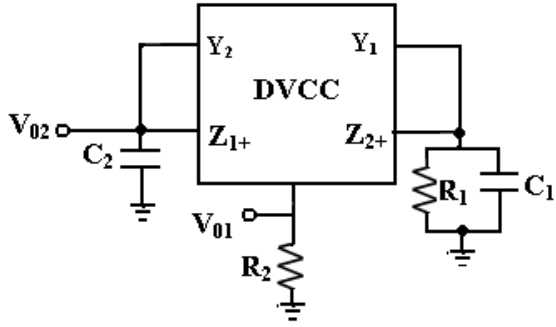

Figure 2. Proposed VMQO

Replacing s=jω and equating real and imaginary terms, the above equation yields the frequency of oscillation (FO) and condition of oscillation (CO) as

$$\text{FO:} \quad \omega_o = \frac{1}{\sqrt{R_1 R_2 C_1 C_2}} \quad (3)$$

$$\text{CO:} \quad \frac{1}{R_1 C_1} + \frac{1}{R_2 C_2} \geq \frac{1}{C_1 R_2} \quad (4)$$

Assuming $2C_1=C_2$; $R_1=2R_2$

$$\omega_o = \frac{1}{R_1 C_1} \quad (5)$$

From Fig. 2, at oscillating frequency, it can be seen that two output voltages are related as

$$V_{01} = j \, V_{02} \quad (6)$$

Thus, the circuit provides two quadrature voltage outputs ($V_{01}$ and $V_{02}$) as shown in the Fig. 3

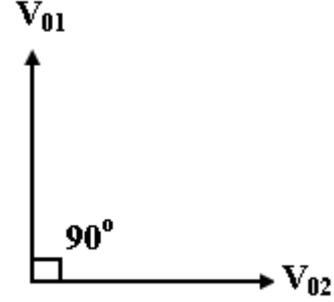

Figure 3. Voltage phasor diagram of VMQO

III. NON-IDEAL ANALYSIS

Taking the tracking errors of the modified DVCC into account, the relationship of the terminal voltages and currents of the DVCC can be rewritten as

$$\begin{bmatrix} V_X \\ I_{Y1} \\ I_{Y2} \\ I_{Z1+} \\ I_{Z2+} \end{bmatrix} = \begin{bmatrix} 0 & \beta_1 & -\beta_2 & 0 \\ 0 & 0 & 0 & 0 \\ 0 & 0 & 0 & 0 \\ \alpha_1 & 0 & 0 & 0 \\ \alpha_2 & 0 & 0 & 0 \end{bmatrix} \begin{bmatrix} I_X \\ V_{Y1} \\ V_{Y2} \\ V_Z \end{bmatrix} \quad (7)$$

where $\alpha_1(s)$, $\alpha_2(s)$ and $\beta_1(s)$, $\beta_2(s)$ represent the frequency transfers of the internal current and voltage followers of the modified DVCC, respectively. If this circuit is working at frequencies much less than the corner frequencies of $\alpha_1(s)$, $\alpha_2(s)$ and $\beta_1(s)$, $\beta_2(s)$, namely, then $\alpha_1(s) = 1-\varepsilon_{i1}$ and $\varepsilon_{i1}$ ($|\varepsilon_{i1}| <<1$) denotes the current tracking error from Z1+ to the X terminal; $\alpha_2(s) = 1-\varepsilon_{i2}$ and $\varepsilon_{i2}$ ($|\varepsilon_{i2}| <<1$) denotes the current tracking error from Z2+ to the X terminal; $\varepsilon_{dki}$ and $\beta_1=1-\varepsilon_{v1}$ and $\varepsilon_{v1}$($|\varepsilon_{v1}| <<1$) denotes the voltage tracking error from the Y1 terminal to the X terminal; $\beta_2=1-\varepsilon_{v2}$ and $\varepsilon_{v2}$($|\varepsilon_{v2}| <<1$) denotes the voltage tracking error from the Y2 terminal to the X terminal of the modified DVCC. The characteristic equation of Fig. 2 becomes

$$s^2 + s\left(\frac{1}{R_1 C_1} + \frac{\beta_2 \alpha_1}{R_2 C_2} - \frac{\beta_1 \alpha_2}{C_1 R_2}\right) + \frac{\beta_2 \alpha_1}{R_1 R_2 C_1 C_2} = 0 \quad (8)$$

The modified frequency of oscillation (FO) and condition of oscillation (CO) are

$$\text{FO:} \quad \omega_o = \sqrt{\frac{\beta_2 \alpha_1}{R_1 R_2 C_1 C_2}} \quad (9)$$

$$\text{CO:} \quad \frac{1}{R_1C_1} + \frac{\beta_2\alpha_1}{R_2C_2} \geq \frac{\beta_1\alpha_2}{C_1R_2} \quad (10)$$

From (9) and (10), the active and passive sensitivities of the VMQO are all low and are obtained as

$$S^{\omega_o}_{\beta_2,\alpha_1} = -S^{\omega_o}_{R_1,R_2,C_1,C_2} = \frac{1}{2}. \quad (11)$$

## IV. SIMULATION RESULTS

The proposed circuit was simulated using PSPICE. The modified DVCC was realized by the CMOS implementation in Fig. 1 [23] using 0.5μm CMOS technology process parameters.

Table 1. Transistor aspect ratios

| MOS transistors | W (μm) | L (μm) |
|---|---|---|
| M1, M2, M3, M4 | 0.8 | 0.5 |
| M5, M6 | 4 | 0.5 |
| M7, M8 | 10 | 0.5 |
| M9, M10, M13 | 14.4 | 0.5 |
| M11, M12, M14 | 45 | 0.5 |

The transistor aspect ratios are listed in Table 1 and the supply voltage used was ± 2.5 V and $V_{BB}$=-1.8 V. The VMQO was designed with $C_1$=10pF, $C_2$=20pF, $R_1$=2KΩ and $R_2$=1KΩ for the oscillation frequency of $f_o$= 7.96MHz. The two quadrature outputs are obtained through PSPICE simulation in voltage forms as shown in Fig. 4. In this figure, the oscillation frequency is 7.86MHz is obtained. The oscillation frequency is 7.86MHz instead of 7.96MHz owing to the effect described in section III. According to (9), this drop-off would be caused by voltage and current tracking error. Fig. 5 shows the simulated Fourier spectrum of $V_{O1}$ and $V_{O2}$. The THD was within 2%, which is low.

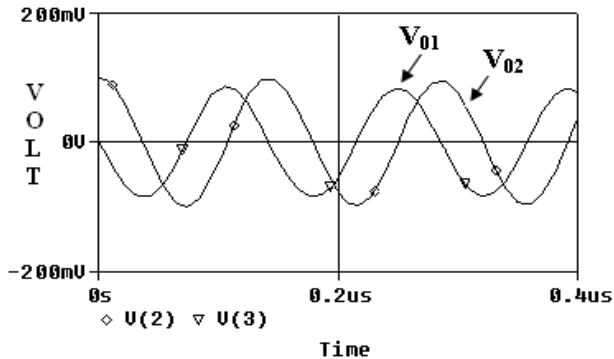

Figure 4. Quadrature voltage outputs of VMQO

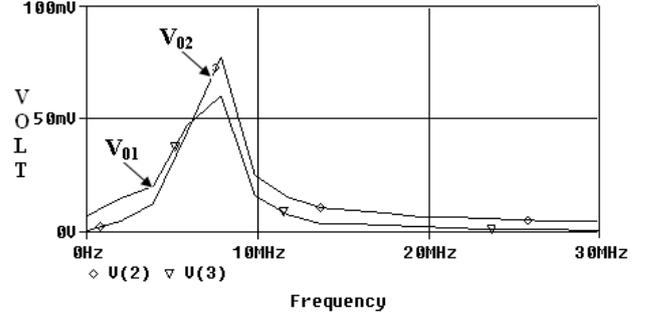

Figure 5. Fourier spectrum of $V_{O1}$ and $V_{O2}$

## V. CONCLUSION

In this paper, a new voltage-mode quadrature oscillator using minimum number of active and passive grounded component is presented. The proposed circuit employs single modified DVCC, two grounded capacitor and two grounded resistors. The use of grounded capacitors and resistors make the proposed circuit suitable for integrated circuit implementation. The active and passive sensitivity are no more than unity. PSPICE simulation results are given to confirm the theoretical analysis. The integration of the proposed circuit is an open area for further research.